\documentclass[pra,reprint,superscriptaddress,longbibliography,nofootinbib]{revtex4-1}

\usepackage{amsmath,amssymb}
\usepackage[euler]{textgreek}

\usepackage{psfrag}
\usepackage{graphicx}
\usepackage{color}
\usepackage[usenames,dvipsnames]{xcolor}
\usepackage{pifont}
\usepackage{fourier}
\usepackage{xr}
\usepackage{stmaryrd}
\usepackage{cancel}

\usepackage{fancyhdr}

\usepackage[process=auto]{pstool}

\usepackage[normalem]{ulem}
\usepackage{euscript,array}

\newcommand{\bea}{\begin{eqnarray}}
\newcommand{\eea}{\end{eqnarray}}
\newcommand{\beq}{\begin{equation}}
\newcommand{\eeq}{\end{equation}}

\definecolor{redg}{rgb}{1,0,0}
\definecolor{blueg}{rgb}{0.22,0.33,0.64}
\definecolor{greeng}{rgb}{0,0.63,0.29}
\definecolor{orangeg}{rgb}{0.96,0.47,0.13}

\DeclareMathAlphabet\mathbfcal{OMS}{cmsy}{b}{n}

\begin{document}

\title{Bosons on a rotating ring with free boundary conditions}

\author{Olindo Corradini}
\affiliation{Dipartimento di Scienze Fisiche, Informatiche e Matematiche, Universit\`a degli Studi di Modena and Reggio Emilia, Via Campi 213/A, I-41125 Modena, Italy,\\ and INFN, Sezione di Bologna, Via Irnerio 46, I-40126 Bologna, Italy}
\author{Antonino Flachi}
\email[]{flachi@phys-h.keio.ac.jp}
\affiliation{Department of Physics \& Research and Education Center for Natural Sciences, Keio University, 4-1-1 Hiyoshi, Kanagawa 223-8521, Japan}
\author{Giacomo Marmorini}
\affiliation{Department of Physics and Mathematics, Aoyama-Gakuin University, Sagamihara, Kanagawa 252-5258, Japan}
\author{Maurizio Muratori}
\affiliation{Dipartimento di Scienze Fisiche, Informatiche e Matematiche, Universit\`a degli Studi di Modena and Reggio Emilia, Via Campi 213/A, I-41125 Modena, Italy,\\ and INFN, Sezione di Bologna, Via Irnerio 46, I-40126 Bologna, Italy}
\author{Vincenzo Vitagliano}
\affiliation{Department of Mathematics and Physics, University of Hull, Kingston upon Hull, HU6 7RX, UK}
\begin{abstract}
We consider a system of interacting non-relativistic bosons confined to a one-dimensional ring in the presence of a synthetic gauge field induced by a rotating barrier. Interactions are introduced as a constraint in field space, and the barrier is modeled by general boundary conditions. Within this setup, we compute the effective action and investigate the profile of the ground state and its sensitivity from rotational velocity and the properties of the barrier. 
\end{abstract}

\maketitle

\newpage

\section{Introduction}

One of the simplest conceivable models in quantum field theory is a free boson confined to a one-dimensional ``box''. This typical textbook example, used to elucidate several basic notions, ranging from second quantization, renormalization, quantum vacuum phenomena, etc., can develop remarkably intricate dynamics once fields are allowed to interact. Equally complex behaviors arise when the underlying background is non-trivial (as in the case of stationary or explicitly time-dependent backgrounds, or in the presence of external potentials). This added level of complexity has made the ``simple'' scalar field setup an active playground for investigating quantum field dynamics in both perturbative and non-perturbative regimes. Novel results have been achieved not only in theoretical aspects but also in concrete experimental settings. Thanks to the advancement in the use of confining optical traps, it is now possible to accurately control the interaction strength of atoms at very low temperatures and the boundary conditions that arise due to the confinement of the particles in space~\cite{Lewenstein:2012,Zohar:2016}.

One-dimensional confined bosons in the presence of a rotating barrier provide a natural (scalar field) setup of the type described above. The complexity in the dynamics manifests then in the form of non-trivial one-loop effects and the occurrence of persistent currents. Usually, persistent currents appear when a quantum field is subject to the influence of a gauge flux with superimposed periodicity, as prescribed by the Aharonov-Bohm effect. A similar phenomenon can occur, although subtly, in stationary setups, as in the presence of rotation, even in the absence of external gauge fields. While rotation does indeed give rise to (artificial) gauge degrees of freedom, the requirement of periodicity rules out the possibility that any physical effect may emerge from such synthetic gauge fields. The obvious reason for this is the gauge invariance that impedes, in such a setting, a laboratory observer to detect any effect of rotation, similarly to a co-rotating observer, making rotation-induced gauge fields \textit{fake}. The situation is different when, in conjunction with rotation, periodicity is modified by physical boundary conditions (in other words, when the barrier rotates - something that can be implemented by the presence of impurities or by a discontinuity, or cut, along the ring). The presence of the barrier breaks the gauge invariance and allows the laboratory observer to detect rotation (differently from what happens for a co-rotating observer that, by definition, does not detect rotation). Then, rotation-induced, \textit{artificial} gauge fields acquire a physical dimension and may lead to the appearance of a persistent current that is in principle observable. Concrete examples of this include the creation of Josephson junctions on a toroidal Bose-Einstein condensates \cite{Ryu:2007}, toroidal Bose-Einstein condensate stirred by a rotating optical barrier \cite{Wright:2013}, or spinor (${}^{87}$Rb) condensates \cite{Beattie:2013}. Ref.~\cite{Cominotti:2014} has thoroughly analyzed such a scenario for an interacting one-dimensional quantum fluid modeled by a non-relativistic scalar field with a delta-function barrier. The analysis of Ref.~\cite{Cominotti:2014} has shown that the presence of the barrier deforms the ideal sawtooth profile of the current to a smeared one, tending to a sinusoid in the strong barrier limit. A corollary of these results with interesting experimental implications for cold atoms on mesoscopic rings is that the current-amplitude reaches an optimal regime, i.e., a maximum, for barriers of any (finite) height, and it is only slightly deformed by the presence of impurities for a large range of interaction strength. 

In the present work, we plan to analyze how quantum effects deform the ground state in the presence of rotation from a different perspective, i.e., within a slightly different setting and following a different approach. First of all, while we will focus on bosons confined on a rotating ring, we will model the effect of the barrier directly through externally-controlled, \textit{free} boundary conditions, rather than adding a delta-function potential with varying strength. While the barrier affects the ground state of the system even in the absence of rotation, a persistent current occurs only when rotation is switched on simultaneously. Here, rather than looking at the current, we will focus on understanding how the ground state is deformed. Secondly, the role of interactions is incorporated in our treatment in a different and somewhat simpler way by imposing a constraint in field space, analogously to what is done in the context of nonlinear sigma models \cite{Zinn-justin}. Finally, our approach will be based on a direct computation of the one-loop effective action and performed using zeta-function regularisation. Extremizing the effective action will give us the properties of the ground state. This approach may also be of interest \textit{per se}. A general discussion of the formalism can be found, for instance, in Ref.~\cite{Toms} for the case of scalar fields in static backgrounds: we will generalize some of those results here to the case of stationary backgrounds.

\section{The model setup}
\label{sec2}

The focus of our work is a non-relativistic complex Scr\"odinger field $\Phi$,
\bea
\Phi = (\phi_1 + i \phi_2)/\sqrt{2},~~~~~~~~~~ \phi_k \in \mathbb{R}
\label{eq2}
\eea
whose dynamics is determined by the action
\bea
S_0 = \int d t \int d x &&\left\{
{i\over 2} \left(\Phi^\dagger \dot{\Phi}- \Phi \dot{\Phi}^\dagger\right)
- {1\over 2m R^2}\left|{\partial \Phi\over \partial \varphi} \right|^2
- V(x)\left|{\Phi} \right|^2 \right.\nonumber\\
&&
\left.+ {i\Omega  \over 2R} \left(\Phi^\dagger {\Phi'} - \Phi {\Phi'}^\dagger\right)
-{m\over 2}\Omega^2 \left|{\Phi} \right|^2 
\right\}.
\label{eq1}
\eea
The above action is obtained from the one at zero rotation $\Omega=0$, after performing a change of coordinates to pass from the  ($\Omega=0$) co-rotating frame to the laboratory frame, $\left(t_0, \varphi_0\right) \rightarrow \left(t,\varphi\right)$, where the index $0$ indicates the coordinates at $\Omega=0$ (we have defined $x=R \varphi$):
\bea
{t} &=&  t_0,~~~~~~~~~~~~~{\varphi} = \varphi_0 + \Omega t_0, 
\label{eq3}
\eea
and
\bea
{\partial \over \partial  t_0} &=& {\partial \over \partial  t}  + \Omega {\partial \over \partial \varphi},~~~~~~~~~~~~~ {\partial \over \partial \varphi_0} = {\partial \over \partial \varphi},
\label{eq5}
\eea
along with the following unitary transformation
\bea
\Phi \to e^{+ i {m\over 2}\Omega^2 t}\Phi. 
\eea
We should first of all notice that rotation in (\ref{eq1}) appears as a constant gauge field, $A_\varphi = m \Omega R$. Furthermore, upon rescaling of the angular velocity in terms of a new angular velocity $\Omega_C$,
\bea
\Omega = {\Omega_C\over m R},
\eea
we see that the above action is the field theoretical equi\-valent of the free part of the Hamiltonian used in Ref.~\cite{Cominotti:2014} (in other words, $\Omega_C$ coincides with the angular velocity of Ref.~\cite{Cominotti:2014}). 
The external potential $V(x)$ can be chosen to incorporate a chemical potential or to be a generic function of the spatial coordinate, as in the case of a confining potential. While it is relatively easy to add an additional external electromagnetic field, we will not explore this possibility here.

\subsection{Normal modes}
Before introducing interactions in the model and proceed with the computation of the effective action and the current, it is instructive to focus on the non-interacting rotating case, for which the calculation can be carried out straightforwardly. In the following we set $V(x)=0$. The equation of motion for the field $\Phi$ (and similarly for $\Phi^\dagger$) is, in absence of any potential,
\bea
i {\partial \Phi \over \partial t} + i {\Omega\over  R} {\partial \Phi\over \partial \varphi} + {1\over \rho}{\partial^2 \Phi \over \partial \varphi^2} - {m\Omega^2\over {2}} \Phi = 0\,,
\label{eom}
\eea
where we have introduced the length scale $\rho$,
\bea
\rho = {2} m R^2.
\eea
Thus, the time-independent Scr\"odinger equation yields the eigenfunctions $f_p(\varphi)$, which satisfy
\bea
{1\over \rho}{\partial^2f_p(\varphi) \over  \partial \varphi^2}
+ i {\Omega\over R} {\partial f_p(\varphi) \over \partial \varphi} 
= \left({1\over 2}m\Omega^2- \lambda_p\right) f_p(\varphi),
\eea
whose solution can be written as
\bea
f_p(\varphi) = N_p e^{-i {\rho\Omega\over 2 R}\varphi} \sin\left(
{\varphi \Delta}
\right),
\eea
where we have defined
\bea
\Delta^2 = \lambda_p  \rho\,,  
\eea
and imposed the condition $f_p(0)=0$. Imposing also the condition $f_p(2 \pi)=0$ gives
\bea
\Delta = {p\over 2},~~~ p \in \mathbb{N}
\eea
from which we can read out the eigenvalues
\bea
\lambda_p = {p^2\over 4\rho}\,.
\label{lambdas}
\eea
Here, we keep the positive values of $p$ in order not to duplicate the solutions. Imposing different boundary conditions, the eigenvalues change, and $p\in \mathbb{Z}$ may have to be included, as in the case of periodic boundary conditions. For completeness, we should mention that, compatibly with the topology of the background, one can require the solution to have an additional dependence on an arbitrary phase $\Xi$, 
\bea
f_p(\varphi) = f_p(\varphi +2 \pi) e^{i\Xi}.
\eea
The above condition, leads to the following constraint for the phase $\Xi$:
\bea
\sin\left({\varphi \Delta }\right)
= e^{-i \left({\rho\Omega\pi\over R} -\Xi\right)} \sin\left(
{(\varphi +2 \pi)\Delta}\right).
\eea
In what follows we set $\Xi=0$.

\subsection{Free fields, Effective action}

The one-loop effective action can be obtained passing to Euclidean time, $t\to - i \tau$, and integrating over quantum fluctuation $\delta \Phi$, where $\Phi = \bar\Phi+\delta \Phi$, and $\bar\Phi$ is a background field. Starting from Eq.~(\ref{eq1}), we obtain the Euclideanized effective action 
\bea
\hspace{-0.1cm}\Gamma &=& \int_0^\beta d\tau \int d x\; 
\bar\Phi^\dagger 
\left[
{\partial \over \partial \tau} 
- {1\over \rho} {\partial^2 \over \partial \varphi^2} - {i\Omega  \over R} {\partial \over \partial \varphi}+{m\over 2}\Omega^2 
\right]\bar\Phi + \delta \Gamma\,,\nonumber
\eea
where
\bea
\delta \Gamma
= \log \det \left(
{\partial \over  \partial \tau} - {1\over \rho}{\partial^2 \over  \partial \varphi^2} - i {\Omega\over R} {\partial  \over \partial \varphi}+{m\over 2}\Omega^2 
\right).
\label{oneloopdet}
\eea
From the above expression, it is evident that the one-loop contribution to the effective action does not depend on the background field $\bar \Phi$. This implies that the effective equations for $\bar\Phi$ will depend only on the background part of the action and not on $\delta \Gamma$. It is important to stress that the background field equation must be equipped with some boundary conditions at the edges of the interval $\left[0,~2\pi\right]$; thus, different solutions for $\bar\Phi$ will arise for different boundary conditions. The interacting case is different, as $\delta \Gamma$ develops a dependence on the background fields.

Assuming periodic boundary conditions in Euclidean time, the complete eigenfunctions will have the form 
\bea
e^{-i \omega_n \tau} f_p(\varphi)\,,
\eea
with the frequencies given by
$$
\omega_n=2 \pi n/\beta,~~~~~~~~~~~~~n\in\mathbb{Z}.
$$ 
The quantity $\beta$ can be thought as the inverse temperature or as the size of the Euclidean box that is let to infinity at the end of the calculations (zero temperature limit). The eigenvalues of the full differential operator in (\ref{oneloopdet}) then become
\bea
{\mathcal E}_{np} = 
i \omega_n 
+ \lambda_p.
\eea


We use zeta-regularization along with the results of the preceding sub-section and express the one-loop effective action in terms of the following \textit{generalized} zeta function,
\bea
\zeta(s) = \sum_{n=-\infty}^\infty \sum_{p} {\mathcal E}_{np}^{-s}\,,
\label{zetasum}
\eea
as
\bea
\delta \Gamma  &=& 2\pi R\beta\left( \zeta(0) \log \ell -  \zeta'(0)\right), 
\label{effacczt}
\eea
with $\ell$ indicating a renormalization scale with dimension of length (see Refs.~\cite{Kirsten,elizalde94} for an introduction to spectral zeta functions and zeta-regularization). The advantage of this approach is that it reduces the problem to the computation of the analytically continued values at $s=0$ of $\zeta(s)$ and its derivative. This is customarily done by finding a (integral) representation for the series (\ref{zetasum}) for which the analytical continuation can be carried out. In the present case, we will limit our discussion to the zero temperature limit, therefore it is convenient to rearrange the zeta function by separating out the zero temperature contribution,
\bea
\zeta(s) = \sum_{p>0} \lambda_p^{-s} + \sigma(s),
\label{zeta0T}
\eea
where the second term,
\bea
\sigma(s) = {\beta^s\over \Gamma(s)} \sum_{n=1}^\infty \sum_{\lambda_p>0} {e^{-n\beta \lambda_p
} \over n^{1-s}},
\eea
encodes the finite temperature corrections. It is easy to show that this term does not contribute to the effective action in the $T\to 0$ limit: 
\bea
\lim_{T\to 0} \sigma'(0) = - \lim_{\beta\to \infty} \sum_p \ln\left(1-e^{-\beta \lambda_p
}\right) = 0.
\eea
Thus, only the first term in (\ref{zeta0T}) contributes to the effective action for vanishing temperature. For $\Omega=0$, the first term corresponds to the zero-point energy or Casimir term. This is nothing but the extension to the stationary case of what can be found in Ref.~\cite{Toms}. In the absence of rotation, this term only produces a constant shift in the effective action, and it does not contribute to the current. In the presence of rotation, $\Omega\neq 0$, in the non-interacting regime, the story is similar. The vacuum energy contribution can be expressed (as it may be expected) in terms of Riemann zeta functions:
it is straightforward to see that the zero-point energy contribution is related to the analytically continued values at $s=0$ of the function
\bea
\xi(s) = (4\rho)^{s}\sum_{p=1}^{\infty} p^{-2s} = (4\rho)^{s} \zeta_R(2s),
\eea
with $\zeta_R(s)$ representing the Riemann zeta. At this point, the analytic continuation is trivial and, following relation (\ref{effacczt}) yields
\bea
\delta \Gamma = 
2 \pi R \beta {1\over 2} \log\left({16\pi^2 \rho\over \ell}\right).
\eea
Some remarks are in order. First of all, we notice that there is no dependence on the angular velocity. This is specific to the (Dirichlet) boundary conditions that we have imposed. Changing boundary conditions to Robin, for instance, will shift the eigenvalues of a quantity depending on $\Omega$, thus reintroducing the angular velocity in the eigenvalues. The analytical continuation in this more general case can be carried out using the Chowla-Selberg formula \cite{ChowlaSelberg,FlachiTanaka}. However, what is more interesting is that imposing Dirichlet boundary conditions leads to an $\Omega$-independent vacuum energy, and therefore implies the vanishing of a persistent current; this is not surprising: Dirichlet boundary conditions correspond to a vanishing flux through the boundary. This agrees with what found in \cite{Cominotti:2014}. 

\section{Interacting non-relativistic problem}

The problem becomes more interesting when interactions are included. The simplest way to incorporate interactions in this model is by enforcing a constraint on the dynamical fields. This can be implemented by a Lagrange multiplier $\lambda( x)$,
\bea
S_\lambda &=& S_0 - \int d t \int d x \lambda( x)\left(\left|\Phi\right|^2 - z^2 \right), 
\label{Slambda}
\eea
where $S_0$ is given by (\ref{eq1}) and $z$ is a constant. Requiring the Lagrange multiplier to extremize the effective action enforces the constraint:
\bea
0 = {\delta S_\lambda \over \delta \lambda}= \left|\Phi\right|^2 - z^2.
\eea
The constraint results in an interaction between the fields, since any change in individual degrees of freedom (e.g., $\phi_1$) is reflected on the other degrees of freedom (e.g., $\phi_2$) that need to change to enforce the constraint, i.e., keep the length $\left|\Phi\right|^2$ constant (and equal to $z^2$).  We should remark that the constrained theory we consider here is different from the \textit{Lieb-Liniger} model considered in \cite{Cominotti:2014}. In our formulation, $z$ represents the coupling constant, and $z \to 0 ~\&~ \lambda \to 0$ returns the noninteracting-dilute limit.

In the interacting case, our first step is also to compute the effective action. Proceeding in a similar way as in the preceding section, we obtain the following expression for the Euclidean effective action to one-loop:
\bea
\Gamma =\! \int_0^\beta\!\!\! d\tau \int\! d x\; 
\bar\Phi^\dagger 
\left[
{\partial \over \partial \tau} 
 - {1\over \rho} \mathcal{D}^2
\right]\bar\Phi
+\lambda( x)\left(\left|\Phi\right|^2 - z^2 \right)
+ \delta \Gamma\,,
\label{effact}~~~~~~~~~
\eea
where 
\bea
\delta \Gamma
= \log \det \left(
{\partial \over  \partial \tau} 
- {1\over \rho} \mathcal{D}^2
+\lambda(x) 
\right),
\label{oneloopdet2}
\eea
and where we have defined the following covariant derivative
\bea
\mathcal{D} = {\partial \over \partial \varphi} +i {\rho \over 2 R} \Omega.
\eea
\subsection{One-loop effective action in the presence of rotation and Lagrange multiplier}
The difference with the non-interacting case lies in the presence of the Lagrange multiplier $\lambda$. Extremization of the effective action with respect to this term controls the dependence on the background field $\bar \Phi$ through the effective field equations. The presence of an \textit{a priori} unknown function in the determinant impedes us to proceed as in the previous section. There are several ways to bypass the problem; here, we will follow an approach (see, for example, Ref.~\cite{NinoTakahiro}) that consists in expressing the effective action in terms of the heat-kernel of the differential operator in (\ref{oneloopdet2}), from which a derivative expansion can be obtained.
Thus, the first step is to re-express the determinant in (\ref{oneloopdet2}) as
\bea
\delta \Gamma
= - \lim_{s\to 0} {d\over ds}
{1\over \Gamma(s)} \sum_{n=-\infty}^{\infty} 
\int_0^\infty {dt \over t^{1-s}} e^{-i\omega_n t} \mbox{Tr}\; e^{- {t\over \rho}D}
\label{oneloopdet3}
\eea
where the differential operator $D$ is defined as
\bea
D = 
-\mathcal{D}^2
+\rho\lambda(x).
\label{Diff}
\eea 
A chemical potential or an external potential can be straightforwardly included in the present treatment. To make the physics more transparent, we proceed by rescaling the integration variable $t = {\rho u}$ leading to
\bea
\delta \Gamma
= - \lim_{s\to 0} {d\over ds}
{ \rho^{s}\over \Gamma(s)} \sum_{n=-\infty}^{\infty} 
 \int_0^\infty {du \over u^{1-s}} e^{-i \varpi_n u} \mbox{Tr}\; e^{-{u} D}\,,~~~~~~~~~
\label{oneloopdet4}
\eea
where 
$$
\varpi_n = {2 \pi n \over \eta},~~~~~~~~~~~~~\eta = {\beta \over \rho}.
$$
Notice that $\eta$ is the ratio of two length scales, thus is dimension-less, and so $\varpi_n$ is. The change of the integration variable essentially corresponds to a rescaling of the inverse temperature $\beta$. Expressing the determinant as in (\ref{oneloopdet4}) has the effect of rescaling the temperature by a factor proportional to $\rho$. This illustrates that the usual small-t (high temperature) heat-kernel asymptotics (see Ref.~\cite{Actor:1987,Toms:1992dq}) occurs in this case for small values of the parameter $\eta$. If we write the functional trace in the above integral in terms of the eigenvalues $\xi_p$ of the operator $D$, we have
\bea
\mbox{Tr}\; e^{-{u} D} = \sum_{\xi_p>0} e^{- \xi_p u}.
\eea
Under the assumption that the eigenvalues are non-negative, the integrand is exponentially suppressed for large-$u$. We will return on the validity of this assumption, and in what follows we will adopt a small-argument approximation for the kernel in the above integral,
\bea
\mbox{Tr}\; e^{-{u} D} \approx K(u) =
{\sqrt{1 \over 2\pi u}} \sum_{k\in \mathbb{N}} a_{k} {u}^{k} + \mbox{boundary terms},~~~~~~~\label{hkexpansion}
\label{hkexp}
\eea
with the coefficients $a_{k}\equiv a_{k}\left(\lambda\right)$ depending on powers and derivatives of $\lambda$. In order to write the bulk equation for the dynamical fields, we only need the bulk part. Boundary contributions will be considered later.
Putting everything together allows us to express the one-loop effective action as
\bea
\delta \Gamma
= - \lim_{s\to 0} {d\over ds}
{ \rho^{s}\over \Gamma(s)} \sum_{n=-\infty}^{\infty} 
 \int_0^\infty {du \over u^{1-s}} e^{-i \varpi_n u} K(u).~~~~~~~~~
\label{oneloopdet5}
\eea
In a region of the complex $s$-plane where the above expression converges, we can swap the summation over $n$ with the integral and re-express the sum using the identity
\bea
\sum_{n=-\infty}^\infty \exp\left(i n u  \right)
=
\eta \sum_{n=-\infty}^\infty \delta\left(u -\eta n\right).
\eea
This allows us to write
\bea
\delta \Gamma
= - \lim_{s\to 0} {d\over ds}
{ \eta \rho^{s}\over \Gamma(s)}  \sum_{n=-\infty}^{\infty} 
 \int_0^\infty {du \over u^{1-s}} \delta\left(u -\eta n\right) K(u)\theta_{reg}(u).~~~~~~~~~
\label{oneloopdet6}
\eea
Formula (\ref{oneloopdet6}) is just a formal re-writing of (\ref{oneloopdet2}) and returns the correct non-interacting ($\lambda\to 0$) limit discussed in the previous section. Also, in order to perform the integration over $u$, we first introduce a regularized step function $\theta_{reg}(u) \to \theta(u)$, that returns the ordinary step function in the limit where the regularization is removed (the details of the regularization are unimportant, as it will become clear shortly). This step is necessary to keep the step function (and the integrand) continuous at $u=0$. Carrying out the integral requires the assumption of continuity at the origin, leading to
\bea
\delta \Gamma
&=&
- \lim_{s\to 0} \lim_{n\to 0}{d\over ds}
{\left({\rho\eta}\right)^{s}\over \Gamma(s)} 
 {n^{s-1}} K\left(\eta n\right) \theta_{reg}(\eta n)\nonumber\\
&&- \lim_{s\to 0} {d\over ds}
{\left({\rho \eta}\right)^{s}\over \Gamma(s)} \sum_{n=1}^{\infty} 
{n^{s-1}} K\left(\eta n\right).
\label{oneloopdet7}
\eea
In the above expression, we have kept the regularized $\theta$-function only for the $n=0$ term in the sum while removed the regularization for the $n>1$ contributions. We may notice that the first term in the heat-kernel expansion does not contribute to the effective action, being independent of $\lambda$ or $\bar\Phi$; we then arrive at
\bea
\delta \Gamma
&=&
-\lim_{n\to 0}\left[{a_0 \over n^{3/2}}+{\eta} {a_1\over \sqrt{n}} \right]{\theta_{reg}(0) \over \sqrt{2\pi \eta}}
\nonumber\\
&&- {1 \over \sqrt{2\pi \eta}} \lim_{s\to 0} {d\over ds}
{\left({\rho \eta}\right)^{s}\over \Gamma(s)} 
\sum_{k\in \mathbb{N}} \zeta_R\left({3\over 2}-{k}-s\right)
\eta^{k}a_{k}\,, \nonumber
\label{oneloopdet8}
\eea
from which we get
\bea
\delta \Gamma
&=&
-\sqrt{\eta\over 2\pi} {a_1}\lim_{n\to 0} {\theta_{reg}(0) \over \sqrt{n}}
%
-
{1\over \sqrt{2\pi\eta}}\sum_{k=1}^\infty \zeta(3/2-k) \eta^k a_k,~~~~~~~~~~
\label{oneloopdet9}
\eea
where we have introduced the renormalization scale $\ell$. We have dropped the term proportional to $a_0$ since it does not depend on the background fields or the Lagrange multiplier and disappears from the equation of motion. Physically, this term corresponds to a renormalization of the vacuum energy. The other term, which diverges in the limit $n \to 0$, proportional to the $a_1$ heat-kernel coefficient, corresponds to a renormalization of the (inverse) coupling $z$ in the classical action. More importantly, we should observe that the expansion appears in powers of $\eta \propto (T m R^2)^{-1}$. {Thus, the expression (\ref{oneloopdet9}) can be safely used in the limit of high temperature and small mass or size, or in the limit of small temperature and large mass or size.} 
We should remark that within this approximation, we are ignoring large-$t$ contributions to the heat-kernel. In principle, these terms should yield infrared-sensitive logs that will repair infrared divergences. It is possible to include such terms using a more elaborate regularization scheme, however here we are simply igno\-ring them. {Based on dimensional analysis, one may conclude that a derivative expansion of the effective action takes the same form (\ref{oneloopdet9}) even beyond the range of validity discussed above.
This, along with the \textit{assumption} that the ground state is not rapidly varying, allows to ignore high-order derivatives. While it is physically reasonable, certainly in a non-relativistic context, to assume that a rapidly varying background is not the ground state, the results of Ref.~\cite{Cominotti:2014} clearly show that this is the case for the present problem.}

The advantage of the present approach lies in the expansion (\ref{hkexpansion}). The coefficients $a_k$ are integrals of local quantities that can be obtained from the knowledge of the differential operator $D$ (See any of the books in Refs~\cite{Avramidi,ParkerToms,Gilkey} for an in-depth introduction). For any operator of the form $\Theta = g^{\mu\nu}\nabla_\mu \nabla_\nu + f(x)$, where $\nabla_\mu$ is any covariant derivative that may include gauge potentials and $f = f(x)$ is any regular function (in general, $f(x)$ is an operator that does not contain derivatives), the coefficients can be found in any of the references \cite{Avramidi,ParkerToms,Gilkey}. In the present case, the metric tensor is trivial, the spin structure absent, and $f(x) \to \lambda(x)$, leading to the following expressions for the first four coefficients that are relevant to our case:
\begin{align}
&a_0= \beta \int dx 1 \nonumber
\\
&a_1=\beta \int dx \left(- \lambda \right)\nonumber
\\
&a_2=\beta \int dx \left(\frac{1}{2}\lambda^2-\frac{1}{6} {\mathcal D}^2 \lambda \right) \nonumber
\\
&a_3=\beta \int dx \left(-\frac{1}{6}\lambda^3 +\frac{1}{12}\left( {\mathcal D}\lambda \right)^2+\frac{1}{6} \lambda{{\mathcal D}^2\lambda}-\frac{1}{60}{{\mathcal D}^4\lambda}\right).
\nonumber 
\end{align}

\subsection{Effective equations and boundary conditions}

Relations (\ref{effact}), (\ref{oneloopdet9}), along with the above explicit form of the coefficients yield an explicit expression for the effective action from which the equation for the background fields, $\bar\Phi$ and its conjugate, and for the Lagrange multiplier $\lambda$ can be obtained. Here, we will truncate the derivative expansion to order $k=3$ (i.e., including up to the coefficient $a_3$), which allows us to obtain the following system of nonlinear coupled differential equations\footnote{The second order equations prior to the first-order reduction are
\bea
0&=& {1\over \rho}{\partial^2 \bar\Phi \over \partial \varphi^2}
+ i {\Omega\over  R} {\partial \bar\Phi\over \partial \varphi} 
- \left({m\Omega^2\over 2} +\lambda(\varphi)\right) \bar\Phi \nonumber\\
0&=&  {d^2 \lambda(\varphi) \over d\varphi^2} -{3} \lambda^2 + {\mathcal M} \lambda 
+ \mathcal{U},
\nonumber
\eea
}:
\bea
X_1' &=& X_2 \\
X_2' &=& {m\rho\Omega^2\over 2} X_1 +\rho Z_1 X_1 + {\rho \Omega\over R}Y_2\\
Y_1' &=& Y_2 \\
Y_2' &=& {m\rho\Omega^2\over 2} Y_1 +\rho Z_1 Y_1 - {\rho \Omega\over R}X_2 \\
Z_1' &=& Z_2 \\
Z_2' &=& 3 Z_1^2 - \mathcal M Z_1 - \mathcal U
\eea
where we have defined ($z_{ren}$ is the renormalized coupling)
\bea
X_1 &=& \Re \bar \Phi,~~~
Y_1 = \Im \bar \Phi,~~~
Z_1 = \lambda,~~~\\
X_2 &=& \Re \bar \Phi',~~~
Y_2 = \Im \bar \Phi',~~~
Z_2 = \lambda',~~~
\eea

\begin{figure*}[t!]
\begin{center}
\begin{tabular}{ccc}
$\Omega=0$&$\Omega=0.5$&$\Omega=1$\\  
\includegraphics[scale=0.42,trim={1cm 3.3cm 1.3cm 3cm},clip]{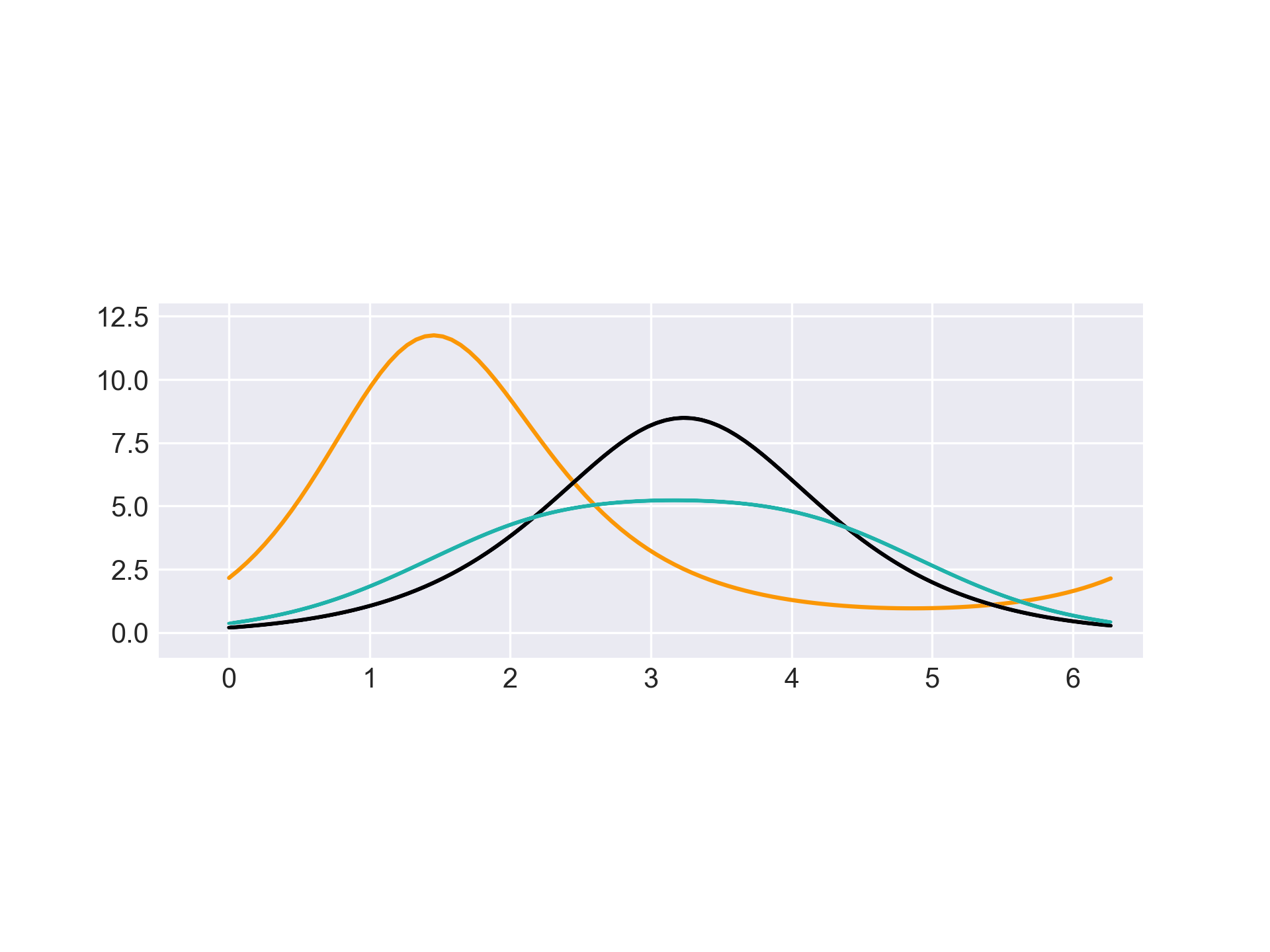}&
\includegraphics[scale=0.42,trim={1.2cm 3.3cm 1.3cm 3cm},clip]{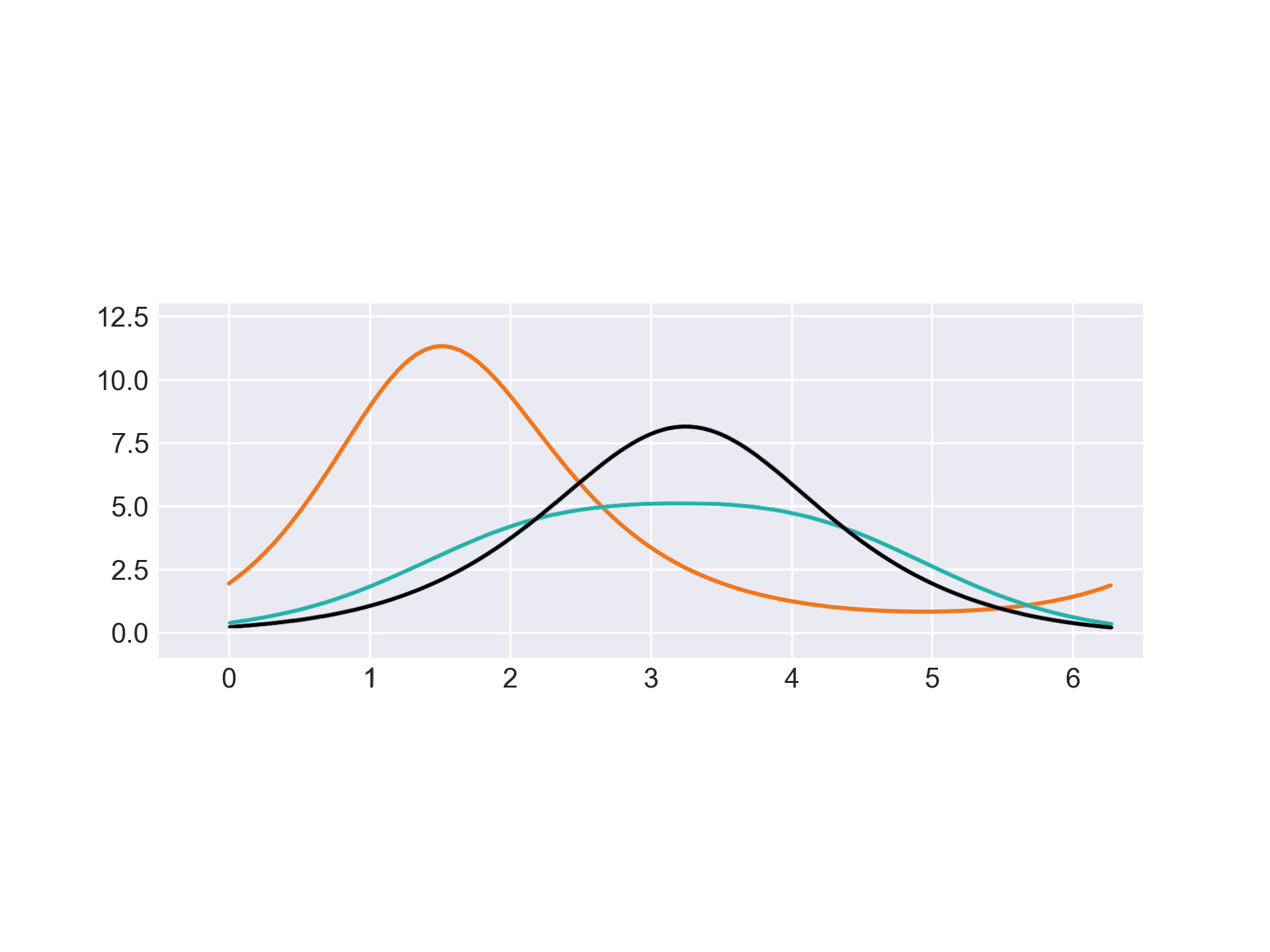}&
\includegraphics[scale=0.42,trim={1.2cm 3.3cm 1cm 3cm},clip]{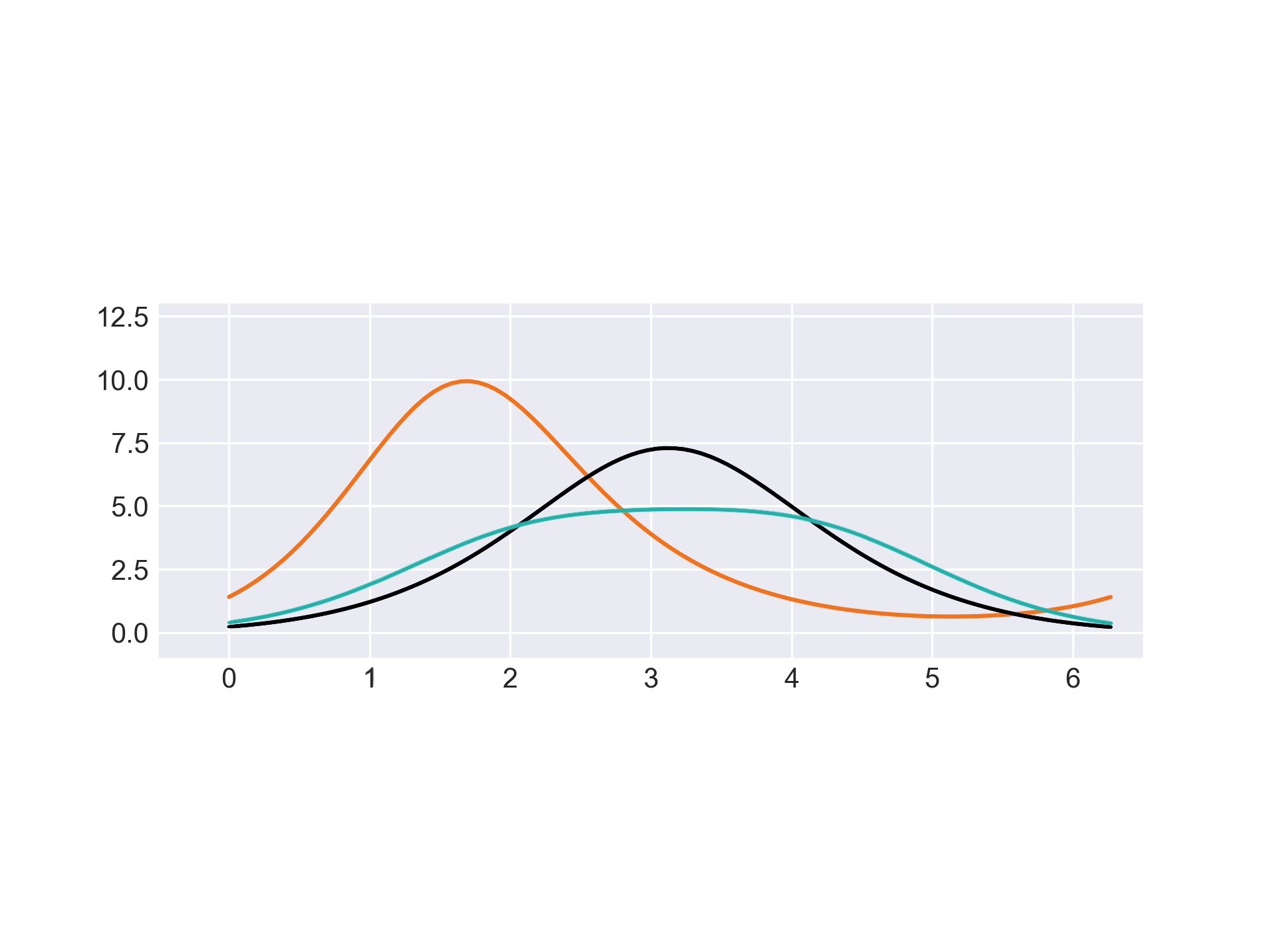}\\
\includegraphics[scale=0.42,trim={1cm 3.5cm 1.3cm 4cm},clip]{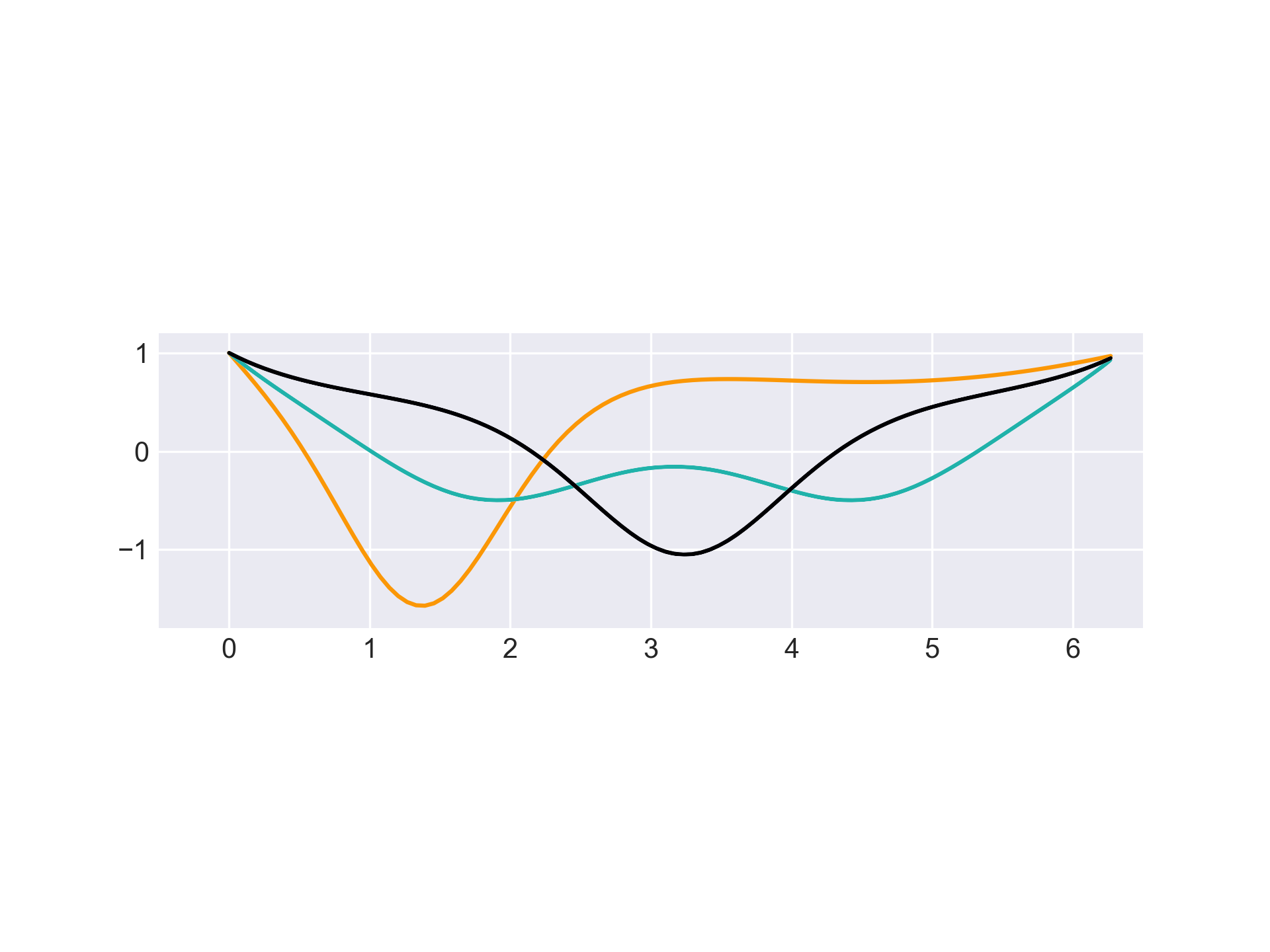}&
\includegraphics[scale=0.42,trim={1.2cm 3.5cm 1.3cm 4cm},clip]{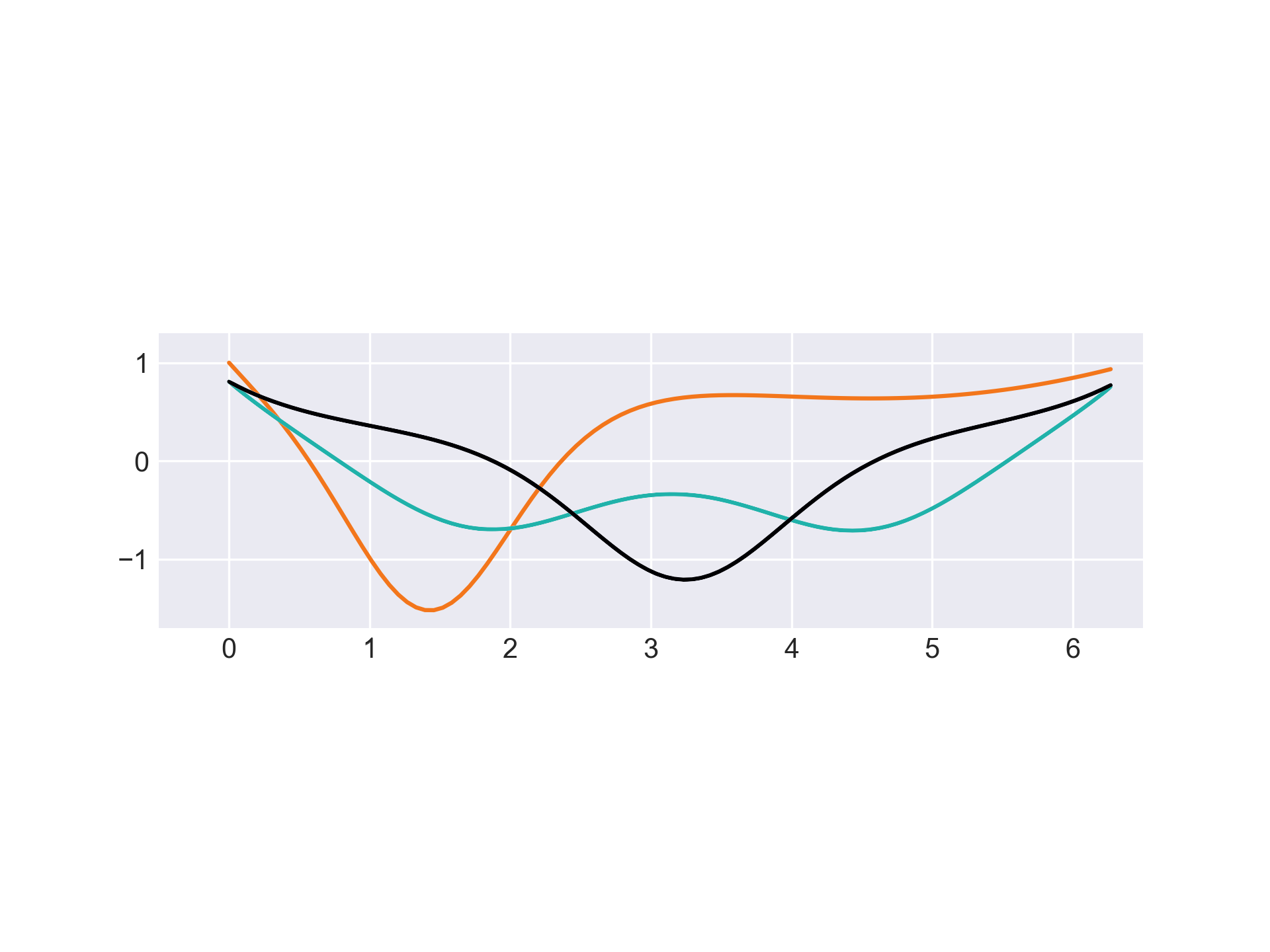}&
\includegraphics[scale=0.42,trim={1.2cm 3.5cm 1cm 4cm},clip]{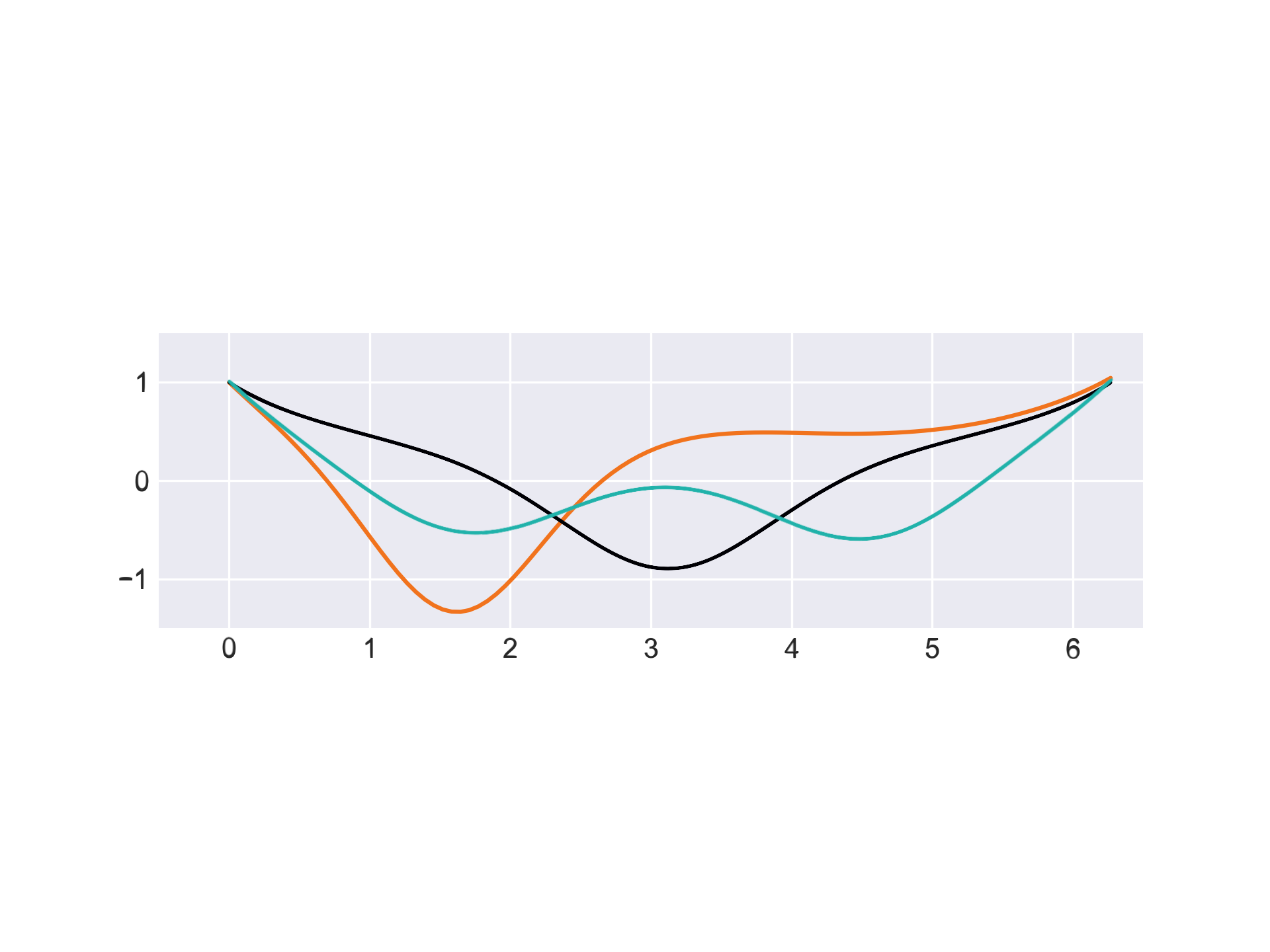}\\
$\varphi$&$\varphi$&$\varphi$
\end{tabular}
 \put(-520,30){$\Phi^2$}
 \put(-513,-30){$\lambda$}
\end{center}
\caption{The figure shows the numerical solutions for $\Phi^2$ and $\lambda$ for illustrative values of the rotational velocity $\Omega$. We have selected the parameters as follows: $m\times R = 0.3$, $\beta/ R= 10$, and $z=0.1$. The curves correspond to the following solutions: for $\Omega=0$, orange $\Rightarrow (\bar \Phi_{\varphi=0}=1.121,~ \bar \Phi'_{\varphi=0}=0.95 , ~ \lambda'_{\varphi=0}=-1.66)$, cyan $\Rightarrow (0.001,0.60,-1.19)$, black $\Rightarrow (0.001,0.44,-0.76)$; 
for $\Omega=0.5$, orange $\Rightarrow (1.011, 0.96, -1.57)$, cyan $\Rightarrow (0.001,0.60,-1.23)$, black $\Rightarrow (0.001,0.44,-0.80)$; 
for $\Omega=1$, orange $\Rightarrow (1.001, 0.58, -1.37)$, cyan $\Rightarrow (0.001,0.61,-1.37)$, black $\Rightarrow (0.001,0.47,-0.92)$.}
\label{figure}
\end{figure*}

\bea
\mathcal U &=&
 {\pi \rho^3 \Omega^2\over 3 R^2 \beta}{\zeta(3/2)\over \zeta(5/2)}
-{6 \rho^2\over \beta^2}{\zeta(1/2)\over \zeta(-3/2)}\nonumber\\
&&
-{6\sqrt{2\pi\rho^5\over \beta^5}}{\left(X_1^2+X_2^2-z_{ren}^2\right)\over \zeta(-3/2)}
-{1\over 10}\left({\rho\Omega\over 2R}\right)^4\,,
\\
\mathcal M &=& \left(8\pi {\zeta(3/2)\over \zeta(5/2)}{\rho\over \beta} 
 - 3\left({\rho \Omega\over 2R}\right)^2\right)\,.
 \eea

As anticipated in the introduction, we are interested in the response of the system (in particular of its ground state) to changes in the boundary conditions. Thus, we enforce the boundary conditions directly on the bulk solutions and see how these will change when the boundary conditions change. We should also notice that since we have introduced interactions as a constraint in field space through a Lagrange multiplier, quantum effects enter in the effective equation for $\lambda$ that, through the nonlinear structure of the effective equations (i.e., the coupling between the background fields $\bar\Phi$ and $\bar\Phi^\dagger$, and $\lambda$), affects the ground state.

The numerical calculation is carried out in \textit{Python} and the equation is solved by fixing the boundary conditions on the left at $\varphi=0$ and shooting to the right. The right-hand boundary is regulated by shifting it by an amount $\epsilon$ increasingly smaller until the solution satisfies the imposed requirements. All plots shown refer to $\epsilon = 10^{-2}$. The boundary values of the real and imaginary parts of $\bar \Phi$  at $\varphi=0$---i.e. $\Re \bar \Phi_{\varphi=0}$ and $\Im \bar \Phi_{\varphi=0}$---and $\lambda_{\varphi=0}$, along with their derivatives have been varied within the following intervals: $10^{-3} \leq \Re \bar \Phi_{\varphi=0} \leq 1.4$, $10^{-3} \leq \Im \bar \Phi_{\varphi=0} \leq 1.4$, $-10^{-3} \leq \Re \bar \Phi'_{\varphi=0} \leq 1.1$, $-10^{-3} \leq \Im \bar \Phi'_{\varphi=0} \leq 1.1$, $-1.90\leq \lambda'_{\varphi=0} \leq 0.81$. Also, we have re-scaled the value of $\lambda$ at $\varphi=0$ to unity. With the boundary values on the right fixed, we have numerically searched for solutions that satisfied continuity and periodicity or anti-periodicity for the real and imaginary part (leading to a periodic modulus square $\Phi^2$) with a tolerance of $1\%$ and repeated the numerical search at increments of $10^{-2}$ on all boundary values. We should note here that  solutions that do not satisfy this added constrains are still valid, despite being non-periodic or discontinuous at the boundary. The values of the physical parameters have been set as follows: $m\times R=0.3$, $\beta/R = 10$, and $z=0.1$. In the numerical simulations we have set $R=1$. {The rationale behind this choice was to keep both the mass and the temperature small. Notice that this choice of parameters requires the additional assumption that the solution is not rapidly varying; in other words, such solutions are eliminated from the spectrum of the possible ones.} Although we do not report them here, we have explored other parameter sets which have led to similar numerical solutions.

Some illustrative results of the numerical calculation are given in Fig.~\ref{figure} for several values of the angular velocity, $\Omega=0, 0.5, 1$. We have also explored the vicinity of each of these values (e.g., for $\Omega=0$, we have checked $\Omega=0.1, 0.2, 0.3$, etc.) without finding any significant deformation in the numerical solutions.

Several remarks are in order. First of all, one should not confuse the variety of solutions with excited states. Each solution corresponds to a specific choice of boundary conditions and is unique; so, despite different solutions leading, in principle, to different values of the action, no transition between them occurs, as long as the boundary conditions are kept fixed. This may be an interesting point, as the boundary conditions can, in principle, be controlled by optical traps. Secondly, some of our solutions clearly show a behavior similar to those of Ref.~\cite{Cominotti:2014}: solutions are peaked at $\varphi=\pi$ and descend smoothly in both directions towards the boundaries. Third, we find \textit{out-of-phase} solutions with larger amplitude, peaked near $\varphi=\pi/2$; such solutions join continuously at the boundaries and are dephased with respect to the solutions peaked at $\varphi=\pi$ (again, these are not higher energy solutions, but just solutions obeying different boundary conditions). It is interesting to notice the similarity between the amount of dephasing and the detuning of the boundary conditions; if boundary conditions deviate from those producing the solutions symmetric with respect to the center of the interval, then the resulting solution acquires a phase. While from the mathematical point of view this is a trivial observation (i.e., it also happens for plane waves), in the present case, it suggests a way to measure a deviation from specified boundary conditions. The numerical profiles of the Lagrange multiplier (which have no correspondence with Ref.~\cite{Cominotti:2014}) correlate with those of the amplitude and show a peak in correspondence to that of the amplitude.

\section{Conclusions}

In this work, we have studied a system of (free and interacting) confined non-relativistic bosons in one dimension in the presence of rotation. 
Confinement in the angular direction occurs due to boundary conditions that can be physically implemented by optical trapping or impurities. Boundary conditions essentially mimic the presence of a barrier and prevent the possibility of gauging away the synthetic gauge field associated with rotation, making the combination rotation-barrier an intriguing way to alter the properties of the system. In this work, we have studied how the ground state (i.e., the extremal of the one-loop effective action) changes when rotation or boundary conditions are changed. After discussing the free case, we have considered the much more complicated problem of interacting fields. Here, we have introduced interactions as a constraint in field space, which slightly simplifies our treatment and the computation of the one-loop effective action. The latter is carried out by using an approach based on heat-kernels. This method, adapted for the stationary case discussed here, allowed us to obtain an expansion in terms of the background fields (and their derivatives), assuming these were generic spatially varying functions (and a particular dimensionless combination of the physical parameters to be small). This approach proves to be rather valuable to deal with the case of general boundary conditions, or in other words, for any barrier's property, that induces an inhomogeneous ground state. Furthermore, although in the numerical calculations we have kept the temperature small, the results include (within the validity of our approximations) also finite temperature effects and can be extended at finite density straightforwardly. The method itself can be a helpful complement to fully non-perturbative numerical calculations.

The machinery developed here has been ultimately implemented numerically, and it allowed us to explore the ground state solution for varying boundary conditions. We have found three classes of solutions, two of which are compatible with the behavior of Ref.~\cite{Cominotti:2014}, presenting a maximum at the center of the interval and symmetrically descending towards the boundaries, where the background field profile attains a minimum. We also found a third type of solution with a similar profile but dephased and with a higher amplitude. Such dephased solutions also reach a minimum, close to $\varphi = 2\pi/3$, and can be mapped into center-symmetric solutions by a translation; that is, such solutions are topologically equivalent. This gives further support to the argument of Ref.~\cite{Cominotti:2014} that the presence and type of impurities only minimally deform the properties of the system (in this case, its ground state). 

\acknowledgements
AF acknowledges the support of the Japanese Society for the Promotion of Science Grant-in-Aid for Scientific Research KAKENHI (Grants n. 18K03626, 21K03540). VV has been partially supported by the H2020 programme and by the Secretary of Universities and Research of the Government of Catalonia through a Marie Sk{\l}odowska-Curie COFUND fellowship -- Beatriu de Pin{\'o}s programme no. 801370. VV would like to thank Laura Bonavera and Joaqu{\'i}n Gonz{\'a}lez-Nuevo for a useful discussion.



\begin{thebibliography}{}

\bibitem{Lewenstein:2012}
M. Lewenstein, A. Sanpera, V. Ahufinger, ``\textit{Ultracold atoms in optical lattices - Simulating quantum many body systems}'', Oxford University Press, Oxford (UK), 2012. 
  


\bibitem{Zohar:2016}
See, for example, E. Zohar, J.I. Cirac, and B. Reznik,
\textit{``Quantum simulations of lattice gauge theories using ultracold atoms in optical lattices'',}
Rep. Prog. Phys. {\bf 79} (2016) 014401.


\bibitem{Ryu:2007}
C. Ryu, P. W. Blackburn, A. A. Blinova, and M. G. Boshier Phys. Rev. Lett. 99  (2007) 260401.

\bibitem{Wright:2013}
K. C. Wright, R. B. Blakestad, C. J. Lobb, W. D. Phillips, and G. K. Campbell, Phys. Rev. Lett. 110 (2013) 025302. 

\bibitem{Beattie:2013}
S. Beattie, S. Moulder, R.J. Fletcher, and Z. Hadzibabic, Phys. Rev. Lett. 110  (2013) 025301.


\bibitem{Cominotti:2014}
M. Cominotti, D. Rossini, et. al, Phys. Rev. Lett. 113  (2014) 025301.


\bibitem{Zinn-justin}
Zinn-Justin ``\textit{Quantum Field Theory and Critical Phenomena}'', Cambridge University Press, Cambridge (UK). 

\bibitem{Kirsten} 
K. Kirsten, ``\textit{Spectral Functions in Mathematics and Physics}'', CRC Press, Boca Raton, (2001)

\bibitem{elizalde94} 
E. Elizalde, S. D. Odintsov, A. Romeo, A. Bytsenko, and S. Zerbini, ``\textit{Zeta  Regularization Techniques with Applications}'', World Scientific, Singapore, (1994)


\bibitem{Toms}
David J. Toms ``\textit{The Schwinger action principle}'', Cambridge University Press, Cambridge (UK). 


\bibitem{ChowlaSelberg}
S. Chowla, A. Selberg, Proc. Nat. Acad. Sci. USA 35 (1949) 371.

\bibitem{FlachiTanaka}
A. Flachi, T. Tanaka, Phys. Rev. {\bf D}78  (2008) 064011.

\bibitem{NinoTakahiro}
A. Flachi, T. Tanaka,
JHEP \textbf{02} (2011) 026.


\bibitem{Actor:1987}
A. Actor, J. Phys. A{\bf 20}, 5351 (1987).

\bibitem{Toms:1992dq} 
  D.~J.~Toms,
  Phys.\ Rev.\ Lett.\  {\bf 69} (1992) 1152;
  Phys.\ Rev.\ D {\bf 47} (1993) 2483.



\bibitem{Avramidi}
I.G. Avramidi,  ``\textit{Heat Kernel and Quantum Gravity}'', Springer, 2000. 


\bibitem{ParkerToms}
L.E. Parker, D.J. Toms, \textit{Quantum Field Theory in Curved Spacetime}, Cambridge University Press (2009).

\bibitem{Gilkey}
P.B. Gilkey,  ``\textit{Invariance Theory, the Heat Equation and the Atiyah-Singer-Index theorem}'', Publish or Perish Inc., USA, 1984. 
  




\end{thebibliography}
\end{document}